\documentclass[showpacs]{revtex4}
\usepackage{amsmath}
\usepackage{amssymb}
\usepackage{graphicx}

\begin{document}

\title{Cosmic string with two interacting scalar fields}
\author{Vladimir Dzhunushaliev
\footnote{Senior Associate of the Abdus Salam ICTP}}
\email{dzhun@krsu.edu.kg} \affiliation{Dept. Phys. and Microel.
Engineer., Kyrgyz-Russian Slavic University, Bishkek, Kievskaya Str.
44, 720021, Kyrgyz Republic}

\author{Vladimir Folomeev}
\email{vfolomeev@mail.ru} \affiliation{Institute of Physics of NAS KR,
265 a, Chui str., Bishkek, 720071,  Kyrgyz Republic}

\author{Kairat Myrzakulov}
\affiliation{Institute of Physics and
Technology, 050032, Almaty, Kazakhstan}

\author{Ratbay Myrzakulov}
\email{cnlpmyra@satsun.sci.kz} \affiliation{Institute of Physics and
Technology, 050032, Almaty, Kazakhstan}

\begin{abstract}
The model of cosmic string formed from two gravitating and interacting scalar fields is considered. It is shown that the regular solutions exist at special choice of the model's parameters only.

\end{abstract}


\maketitle

\section{Introduction}
The idea about an opportunity of existence in the present Universe of large-scale topological defects in the form of cosmic strings was popular in the 1980's and early 1990's~\cite{Bran}. It was called up for explanation of formation of the large-scale structure of the Universe in postinflationary period~\cite{Bran1,Linde}. At a later time, an existence of cosmic strings was called into question because of the obtained observations about CMB anisotropy and pulsar
timing. However, in view of the certain successes in creation of superstring theories (in particular, M-theory),  new interest for creation of models of such objects has arisen.

Cosmic strings may be formed in the early Universe at phase transitions with spontaneous symmetry breaking~\cite{Bran}. Topologically stable strings do not have ends: they can be either infinitely long or form closed loops. The simplest model of cosmic strings is a model with a complex scalar field $\phi$. In this case the complex field can be divided into a
pair of two real scalar fields: $\phi = \phi_1+i \phi_2$. As a potential energy of such field, the well-known "Mexican hat" potential can be chosen:
\begin{equation}
\label{pot1}
V(\phi_1,\phi_2)=\frac{\lambda}{4}(|\phi|^2-\eta^2)^2=\frac{\lambda}{4}(\phi_1^2+\phi_2^2-\eta^2)^2,
\end{equation}
where $\lambda$ is a self-coupling constant, and $\eta$  the scale of symmetry breaking. It is obvious that such potential possesses phase symmetry, i.e. it remains invariant under the U(1)
group of global phase transformations $\phi(x) \rightarrow e^{i\alpha}\phi(x)$. Such
a potential has a minimum not in zero but on a circle with fixed radius $|\phi| = \eta$ which is a degenerate ground state. At high temperatures, presence of the central hump is not essential due to large fluctuations of the scalar field. However, as the temperature falls down, the field tends to one of the ground states that, accordingly, leads to
symmetry breaking. It appears that in this case an origin of the linear defects - cosmic strings - is possible. This process is similar to the process of occurrence of defects in  condensed matter under the low-temperature phase transformations (Abrikosov filaments in the theory of superconductivity~\cite{Abrik}, magnetic flux tubes in superconductors, etc.).

There are many works devoted to examination of topological defects in different aspects (see, e.g., \cite{Bran2}
and references therein). One of the direction consists in consideration of models of cosmic strings with use of
various types of scalar fields~\cite{Baze}, including the models with two interacting
scalar fields~\cite{BezerradeMello:2003ei}. Another possibility consists in consideration of models of cosmic strings formed from classical
Yang-Mills fields. For example, it is shown in Ref.~\cite{Gal'tsov:2006tp} that such strings within the framework of the SU(2) theory really exist.

In this Letter we offer the model of cosmic string using the special form of the potential energy of scalar field:
\begin{equation}
\label{pot2}
V(\phi,\chi)=\frac{\lambda_1}{4}(\phi^2-m_1^2)^2+\frac{\lambda_2}{4}(\chi^2-m_2^2)^2+\phi^2 \chi^2-V_0.
\end{equation}
Here $\phi, \chi$ - two real scalar fields with masses $m_1, m_2$ and the self-coupling constant $\lambda_1, \lambda_2$ accordingly, $V_0$ is a normalization constant. This potential was obtained in Ref. \cite{Dzhunushaliev:2006di} at approximate modeling of a condensate of
gauge field in the SU(3) Yang-Mills theory. Unlike the potential (\ref{pot1}), the potential (\ref{pot2}) has two global
minimums at $\phi=0, \chi = \pm m_2$ and two local minimums at $\chi=0, \phi = \pm m_1$ at
values of the parameters $\lambda_1, \lambda_2$ used in this
paper. The conditions for existence of
local minima are: $\lambda_1>0, m_1^2>\lambda_2 m_2^2/2$, and for
global minima: $\lambda_2>0, m_2^2>\lambda_1 m_1^2/2$. From
comparison of the values of the potential in global and local minima
we have (taking into account that $V_{loc}=0$ at $V_0=\lambda_2 m_2^4/4$, see below):
$V_{gl}<V_{loc} \Longrightarrow V_{gl}<0$, i.e. $\lambda_2
m_2^4> \lambda_1 m_1^4$. Besides two local and two global minima,
four unstable saddle points exist:
$$
\phi=\pm \sqrt{\frac{\lambda_2}{2}}\sqrt{m_2^2-\frac{\lambda_1
\lambda_2 m_2^2-2 \lambda_1 m_1^2}{\lambda_1 \lambda_2-4}}, \quad
\chi=\pm \sqrt{\frac{\lambda_1
\lambda_2 m_2^2-2 \lambda_1 m_1^2}{\lambda_1 \lambda_2-4}}.
$$

As is known from the previous examinations of the given potential for the spherically symmetric case (boson star)~\cite{Dzhun}, solutions with finite energy can exist only at the
particular values of the parameters $m_1, m_2$. In this sense, the problem is reduced to finding of eigenfunctions $\phi, \chi$ and eigenvalues of the specified parameters. As it was shown in Ref.~\cite{Dzhun}, the solutions in this case tend to the local minimum.
Let us note that the very similar potential was considered in Ref.~\cite{BezerradeMello:2003ei}

\section{Equations and solutions}

The specified features of behaviour of the solutions allow to hope for an opportunity of obtaining of regular solutions at modeling  of cosmic strings with the given form of potential. For that we will use the static cylindrical metric~\cite{Fj}:
\begin{equation}
\label{metr}
ds^2=c^2 e^{2\nu}dt^2-e^{2(\gamma-\psi)}dr^2-e^{2\psi}dz^2-r^2e^{-2\psi}d\theta^2,
\end{equation}
where $t\in ]-\infty, \infty[$, $r\in [0, \infty]$, $z\in ]-\infty, \infty[$, $\theta \in [0,2\pi[$,
and the functions $\nu, \psi, \gamma$ depend only from $r$. The metric describes an
infinite string where the source of gravitation is scalar fields $\phi$ and $\chi$ with the Lagrangian:
\begin{equation}
\label{lagr}
L=\frac{1}{2}\partial_\mu \phi \partial^\mu
\phi+\frac{1}{2}\partial_\mu \chi \partial^\mu
\chi-V(\phi,\chi)
\end{equation}
with the potential energy $V(\phi, \chi)$ from (\ref{pot2}). Then the system of gravitational and field equations (in geometrical units $c=1, 8\pi G=1$) will be:
\begin{eqnarray}
\label{sys}
\psi^{\prime \prime}+\frac{1}{r}\psi^\prime-2 r e^{2(\gamma-\psi)} V \psi^\prime&=&-e^{2(\gamma-\psi)} V,\nonumber \\
\nu^{\prime \prime}+\frac{1}{r}\nu^\prime-2 r e^{2(\gamma-\psi)} V \nu^\prime&=&-e^{2(\gamma-\psi)} V,\nonumber \\
\frac{\gamma^\prime-\nu^\prime-\psi^\prime}{r}&=&2 e^{2(\gamma-\psi)} V,\\
\phi^{\prime \prime}+\frac{1}{r}\left[1+r(\psi^\prime+\nu^\prime-\gamma^\prime)\right]\phi^\prime&=&e^{2(\gamma-\psi)}\phi
\left[2\chi^2+\lambda_1(\phi^2-m_1^2)\right], \nonumber\\
\chi^{\prime \prime}+\frac{1}{r}\left[1+r(\psi^\prime+\nu^\prime-\gamma^\prime)\right]\chi^\prime&=&e^{2(\gamma-\psi)}\chi
\left[2\phi^2+\lambda_2(\chi^2-m_2^2)\right].\nonumber
\end{eqnarray}
(The first equation is obtained from the following combination of the components
of the Einstein equations: $(_\theta^\theta)-(_z^z)+(_r^r)+(_t^t)$, the second one - from: $(_\theta^\theta)+(_z^z)+(_r^r)-(_t^t)$, and the third one - from: $(_t^t) - (_r^r)$).
\begin{figure}[h]
\begin{minipage}[t]{.5\linewidth}
  \begin{center}
  \fbox{
  \includegraphics[height=6cm,width=8.4cm]{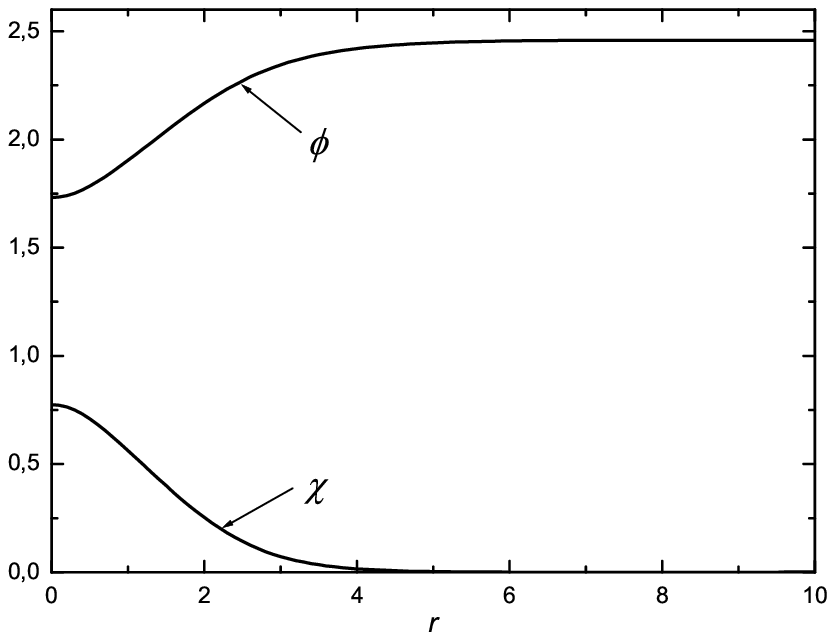}}
  \caption{The functions $\phi, \chi$.}
    \label{phch}
  \end{center}
\end{minipage}\hfill
\begin{minipage}[t]{.5\linewidth}
  \begin{center}
  \fbox{
  \includegraphics[height=6cm,width=8.4cm]{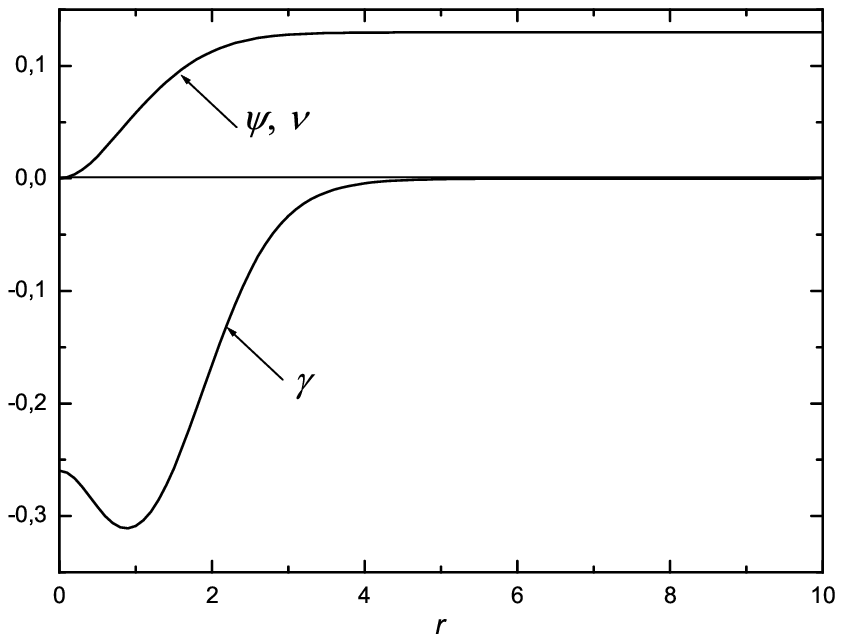}}
  \caption{The functions $\psi, \nu, \gamma$.}
  \label{met}
  \end{center}
\end{minipage}\hfill
\end{figure}

As it was noted above, the given system has the regular solutions only at a special choice of the parameters $m_1, m_2$. The procedure of finding of such solutions is described in Ref.~\cite{Dzhun}. Its essence is reduced to the following: setting some boundary
conditions, solutions of the system (\ref{sys}) are searching by a shooting method. For this purpose, initial values of the parameters $m_1, m_2$ are choosing, and the behaviour of the solutions is explored. At arbitrary choice of $m_1, m_2$, the solutions are divergent either
in $+\infty$ or in $-\infty$. Sequentially carrying out a series of iterations, it is possible to find such values of $m_1, m_2$ at which the solutions are regular. Thus the problem is reduced to finding of the eigenvalues of the parameters $m_1, m_2$ for the nonlinear system (\ref{sys}).

As an example, the solutions for (\ref{sys}) are presented in Figs. (\ref{phch},\ref{met}) at
the following boundary conditions:
\begin{alignat}{2}
\label{in}
\phi(0)&=\sqrt{3},&  \phi^\prime(0)&=0, \nonumber \\
\chi(0)&=\sqrt{0.6},& \chi^\prime(0)&=0, \nonumber \\
\psi(0)&=0,&  \psi^\prime(0)&=0,\\
\nu(0)&=0,& \nu^\prime(0)&=0,\nonumber \\
\gamma(0)&=-0.2597281.& \nonumber
\end{alignat}
The value of $\gamma (0)$ is chosen with the purpose of obtaining an asymptotically zero value of the function $\gamma$. As evident from the figure, the functions $\psi, \nu$ tend
to the constant value asymptotically. With the purpose of obtaining an asymptotically
cylindrical Minkowski metric, these functions can be renormalized by redefinition of the variables $t, r$ and $z$ as follows:
$t\rightarrow t e^{-\nu_\infty}, r\rightarrow r e^{\psi_\infty}, z\rightarrow z e^{-\psi_\infty}$, where $\nu_\infty, \psi_\infty$ are asymptotical values of the functions $\nu$ and $\psi$ accordingly.

At the specified boundary conditions and the choice of the parameters $\lambda_1=0.1, \lambda_2=1$, the following values of the parameters have been found:  $m_1 \approx 2.4600781$ and $m_2 \approx 3.01777633$. At these values $\phi \rightarrow m_1$ and $\chi \rightarrow 0$
that corresponds to tending of the solution to the local minimum of the potential (\ref{pot2}). Here it turns out to be zero that was provided by corresponding choice of the value of $V_0=(\lambda_2/4) m_2^4$.

\begin{figure}[h]
\begin{center}
\fbox{
  \includegraphics[height=6cm,width=8.4cm]{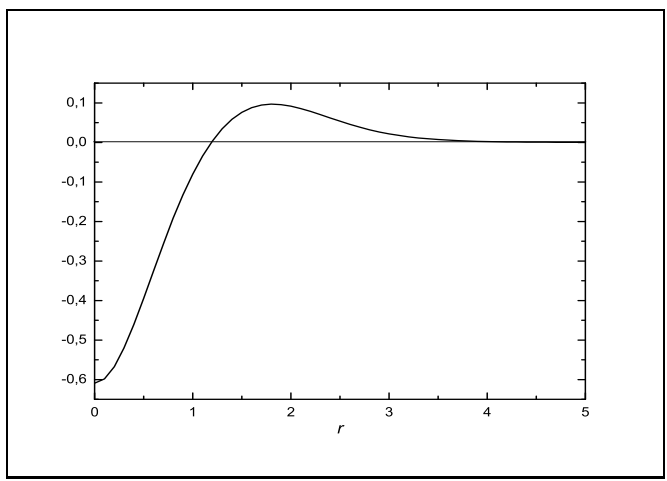}}
 \caption{The energy density $\varepsilon(r)$.}
\label{energ1}
\end{center}
\end{figure}

Let us estimate the mass per unit length $\mu$ of the cosmic string. For this purpose, let us
calculate the energy density:
\begin{equation}
\label{energ}
\varepsilon(r)=\frac{1}{2}e^{-2(\gamma-\psi)}\left(\phi^{\prime 2}+\chi^{\prime 2}\right)+V,
\end{equation}
which is presented in Fig. (\ref{energ1}). The corresponding linear mass will be:
\begin{equation}
\mu=\sqrt{\frac{\pi}{2}}\int_{0}^{\infty}\varepsilon(r) r dr=0.145157\, m_{pl}/l_{pl}\approx 10^{27} {\textrm{g/cm}}.
\end{equation}
This value differs significantly from the typical value
$\mu\approx 10^{21} {\textrm{g/cm}}$~\cite{Bran}.

For analysis of possibility of obtaining more acceptable values of the mass, let us find
asymptotic solutions for the system (\ref{sys}). We will seek for the solutions in the form:
$$
\phi = m_1-\delta \phi, \quad \chi=\delta \chi, \quad \psi=\psi_{\infty}-\delta \psi, \quad
\gamma=\delta \gamma.
$$
Here $\delta \phi, \delta \chi, \delta \psi, \delta \gamma \ll 1$, and here and further subscript ``$\infty$'' means
asymptotic value of the variable. In this context we will have the following asymptotic equations (we omitted the
equation for $\nu$ taking into account that $\nu=\psi$):
\begin{eqnarray}
\label{asympt}
\delta \psi^{\prime \prime} - \frac{1}{2 r} \delta \gamma^{\prime}&=&0, \nonumber \\
\delta \gamma^{\prime} +2 \, \delta \psi^{\prime}&=&2
e^{-2\psi_{\infty}}\left[ \lambda_1 m_1^2 \delta \phi^2+
(m_1^2-2 m_2^2)\, \delta \chi^2\right], \nonumber \\
\delta \phi^{\prime \prime} +\frac{1}{r} \delta \phi^{\prime}&=&2e^{-2\psi_{\infty}}\lambda_1 m_1^2 \delta \phi, \\
\delta \chi^{\prime \prime} +\frac{1}{r} \delta \chi^{\prime}&=&
e^{-2\psi_{\infty}}(2 m_1^2-\lambda_2 m_2^2) \, \delta \chi.
\nonumber
\end{eqnarray}
with asymptotical solutions:
\begin{eqnarray}
    \delta \phi &\approx& \phi_\infty
    \frac{\exp{\left(- e^{-\psi_{\infty}}\sqrt{2\lambda_1 m_1^2} \,\, r \right)}}{\sqrt{r}},
\label{sol1}\\
    \delta \chi&\approx&\chi_{\infty}
    \frac{\exp{\left(- e^{-\psi_{\infty}} \sqrt{2 m_1^2-\lambda_2 m_2^2} \,\, r  \right)}}{\sqrt{r}},
\label{sol2}\\
\delta \psi &\approx& \frac{1}{8}\,\, \phi_{\infty}^2
\frac{\exp{\left( -2 e^{-\psi_{\infty}}\sqrt{2 \lambda_1 m_1^2}
\,\, r \right)}}{r}+ \frac{1}{4} \chi_{\infty}^2\left(
\frac{m_1^2-2 m_2^2}{2 m_1^2-\lambda_2
m_2^2}\right)\frac{\exp{\left( -2 e^{-\psi_{\infty}} \sqrt{2
m_1^2-\lambda_2 m_2^2} \,\,r\right)}}{r},
\label{sol3} \\
\delta \gamma &\approx& -\frac{1}{2}\,\, \phi_{\infty}^2
e^{-\psi_{\infty}}\sqrt{2 \lambda_1 m_1^2}\,
\exp{\left( -2 e^{-\psi_{\infty}} \sqrt{2 \lambda_1 m_1^2} \,\,r\right)}- \nonumber \\
&&  \qquad \qquad \qquad \qquad \qquad \qquad \qquad
\chi_{\infty}^2 \frac{m_1^2-2 m_2^2}{\sqrt{2 m_1^2-\lambda_2
m_2^2}} e^{-\psi_{\infty}} \exp{\left( -2 e^{-\psi_{\infty}}
\sqrt{2 m_1^2-\lambda_2 m_2^2} \,\,r\right)}. \label{sol4}
\end{eqnarray}
In our case, we can neglect the second terms in the right side of
Eqs. (\ref{sol3}, \ref{sol4}) in comparison with the first ones.
Therefore we will have the following expressions for $\delta \psi$
and $\delta \gamma$:
\begin{eqnarray}
\delta \psi &\approx& \frac{1}{8}\,\, \phi_{\infty}^2
\frac{\exp{\left( -2 e^{-\psi_{\infty}}\sqrt{2 \lambda_1 m_1^2} \,\, r \right)}}{r},\\
\delta \gamma &\approx& -\frac{1}{2}\,\, \phi_{\infty}^2
e^{-\psi_{\infty}}\sqrt{2 \lambda_1 m_1^2} \exp{\left(- 2
e^{-\psi_{\infty}} \sqrt{2 \lambda_1 m_1^2}\,\, r \right)}.
\end{eqnarray}

For reducing of the linear mass density, it is necessary to increase the diameter of the
string. As one can see from the solutions, for this purpose one should decrease the parameters $\lambda_1, m_1$
keeping the relation $m_1^2>\lambda_2 m_2^2/2$. By varying of the parameters, one can reduce the value of the mass slightly.
But, as it follows from calculations,
the solution is very sensitive to eigenvalues of the parameters $m_1, m_2$.
Measure of inaccuracy of the numerical method (Runge-Kutta) inhibits performing calculations
for $m_1, m_2$ with sufficient accuracy. Apparently, these calculations could be done using
another numerical methods for solution of differential equations.
So the question about obtaining necessary value of the linear mass density demands further calculations.

Summarizing, we have presented the model of cosmic string created by two interacting scalar fields.
The obtained regular solution exists in a case when the scalar fields tend to the local minimum at infinity.
It is possible that these fields describe a non-perturbative quantized SU(3) gauge field~\cite{Dzhunushaliev:2006di}.
In this sense, our solution could be considered as a quantum analog of the string from Ref.~\cite{Gal'tsov:2006tp}.

We understand that the obtained solution is not a cosmic string in usual sense,
since it tends to false vacuum asymptotically at $V=V_{loc}$. As regards the possibility of
existence of solutions with true vacuum at $V=V_{gl}$, then our numerical calculations
show absence of such solutions. At least for the case without gravitation, there is the Derrick's theorem~\cite{Derr}
 that forbids existence of static solutions for scalar
fields with finite energy for spaces with dimensionality $D>2$.

The authors would like to thank the referee who suggested several proposals improving the paper.

\end{document}